\newcommand{\avg}[1]{\left< #1 \right>} 

\newcommand*{\rev}[1]{}
\newcommand*\diff{\mathop{}\!\mathrm{d}}
\documentclass[manuscript]{aastex6}
\usepackage{amsbsy}
\usepackage{amsfonts}
\usepackage{amssymb}
\usepackage{graphicx}
\usepackage{graphics}
\usepackage{bm}

\begin{document}
\title{Density Fluctuations in a Polar Coronal Hole}
\author{Michael Hahn\altaffilmark{1}, Elke D'Huys\altaffilmark{2}, and Daniel Wolf Savin\altaffilmark{1}}
\altaffiltext{1}{Columbia Astrophysics Laboratory, Columbia University, 550 West 120th Street, New York, NY 10027 USA} \email{mhahn@astro.columbia.edu}
\altaffiltext{2}{Royal Observatory of Belgium, Solar-Terrestrial Centre of Excellence, Ringlaan -3- Av. Circulaire, 1180 Brussels, Belgium}

\date{\today}
\begin{abstract}

We have measured the root-mean-square (RMS) amplitude of intensity fluctuations, $\Delta I$, in plume and interplume regions of a polar coronal hole. These intensity fluctuations correspond to density fluctuations. Using data from the \rev{Sun Watcher using Active Pixel System detector and Image Processing} (SWAP) on \textit{Project for Onboard Autonomy (Proba2)}, our results extend up to a height of about 1.35~$R_{\sun}$. One advantage of the RMS analysis is that it does not rely on a detailed evaluation of the power spectrum, which is limited by noise levels to low heights in the corona. The RMS approach can be performed up to larger heights where the noise level is greater, provided that the noise itself can be quantified. At low heights, both the absolute $\Delta I$, and the amplitude relative to the mean intensity, $\Delta I/I$, decrease with height. However, starting at about 1.2~$R_{\sun}$ $\Delta I/I$ increases, reaching 20--40\% by 1.35~$R_{\sun}$. This corresponds to density fluctuations of $\Delta n_{\mathrm{e}}/n_{\mathrm{e}} \approx$~10--20\%. The increasing relative amplitude implies that the density fluctuations are generated in the corona itself. One possibility is that the density fluctuations are generated by an instability of Alfv\'en waves. This generation mechanism is consistent with some theoretical models and with observations of Alfv\'en wave amplitudes in coronal holes. Although we find that the energy of the observed density fluctuations is small, these fluctuations are likely to play an important indirect role in coronal heating by promoting the reflection of Alfv\'en waves and driving turbulence.

\end{abstract}

	
\maketitle
	
\section{Introduction}\label{sec:intro}

Coronal holes are regions where the magnetic field of the Sun is open. They are the source of the fast solar wind \citep{Zirker1977}. One of the major theories to explain the heating of these regions is based on Alfv\'en wave turbulence \citep[e.g.,][]{Suzuki2006, Hollweg2007, Cranmer2007}. Alfv\'en waves are observed in coronal holes, where they are believed to be excited at the base of the corona and travel outward along the open field lines. Reflected waves are expected to be generated by gradients in the Alfv\'en speed. These reflected waves then propagate inward, interact nonlinearly with the outward propagating waves and thereby drive turbulence and heating. Recent observations suggest that Alfv\'en waves do indeed dissipate at low heights in coronal holes \citep{Bemporad:ApJ:2012, Hahn2012, Hahn:ApJ:2013}. 

For this wave heating model to be viable, waves must be reflected efficiently enough to reach the heating rates required to heat the corona and accelerate the fast solar wind. A problem for the model is that the large-scale gradients in coronal holes are not steep enough to cause sufficient reflection. One possible resolution to this problem is that Alfv\'en wave reflection is enhanced by small-scale density fluctuations along the field lines. Recent calculations have shown that such fluctuations can, in principle, significantly increase the rate of Alfv\'en wave reflection, dissipation, and heating \citep{vanB2016, vanB2017}. 

In order to refine these models, more detailed information is needed characterizing the density fluctuations. Density fluctuations can be studied through observations of emission line intensity oscillations. For most extreme ultraviolet (EUV) emission lines of interest, the intensity is proportional to the square of the electron density, $I \propto n_{\mathrm{e}}^2$. Thus, density fluctuations are related to intensity fluctuations as  $\Delta I/I \propto 2 \Delta n_{\mathrm{e}} / n_{\mathrm{e}}$. 

There have been a number of previous studies of density fluctuations to determine their amplitudes and frequencies, however most were studied using on-disk data or limited to low heights $\lesssim 1.05$~$R_{\sun}$. These works found amplitudes to be in the range of $\Delta n_{\mathrm{e}} /n_{\mathrm{e}} \approx 5$--10\% \citep{DeForest:ApJ:1998, Popescu:AA:2005, McIntosh:AA:2010}. Alfv\'en waves are observed to large heights and so the height range above 1.05~$R_{\sun}$ is important for modeling Alfv\'en wave reflection, but there are few observations of density fluctuations at these heights.



At heights $\gtrsim 1.4$~$R_{\sun}$, there have been studies using the Ultraviolet Coronagraph Spectrometer (UVCS) on the \textit{Solar and Heliospheric Observatory (SOHO)}. Using an O~\textsc{vi} line at 1.4~$R_{\sun}$, \citet{Mancuso:AA:2016} found intensity oscillations with periods of $\sim 20$~minutes and amplitudes of $\Delta I/I \approx 10\%$ ($\Delta n_{\mathrm{e}}/ n_{\mathrm{e}} \approx 5\%$). It is interesting that these large amplitudes, similar to those found near the limb, were found at such large heights as dissipation is expected to reduce acoustic wave amplitudes as the waves propagate \citep{Ofman:ApJ:1999, Ofman:ApJ:2000}. At still larger heights, from 1.5--6~$R_{\sun}$, \citet{Miyamoto:ApJ:2014} used radio occultation measurements to observe compressive waves with amplitudes growing from about $\approx 1\%$ to $\approx 30\%$ over their observed height range. The relevance of that work, though, to the present study is limited as those observations took place during a period of high solar activity and do not appear to have observed a coronal hole.

Here, we study the amplitudes of intensity fluctuations at intermediate heights in a coronal hole from about 1--1.35~$R_{\sun}$ using data from the \rev{Sun Watcher using Active Pixel System detector and Image Processing} (SWAP) on the \textit{Project for Onboard Autonomy (Proba2)}. The advantage of SWAP is its wide field of view of $54^{\prime} \times 54^{\prime}$, allowing observations of the corona out to nearly 2~$R_{\sun}$. The effective maximum height, though, is less than this due to the low signal at large heights.

The rest of this paper is organized as follows: In Section~\ref{sec:obs} we describe the instrument and observations, then we discuss the analysis of the intensity fluctuations in Section~\ref{sec:rms} including a detailed discussion of noise sources. Section~\ref{sec:discuss} discusses the periods of the fluctuations, estimates the energy flux required to sustain the density fluctuations, and describes a possible explanation for the excitation of the density fluctuations as well as the implications for theories of coronal heating. Section~\ref{sec:conclusions} concludes.


\section{Instrument and Observations}\label{sec:obs}

\begin{figure}
	\centering \includegraphics[width=0.9\textwidth]{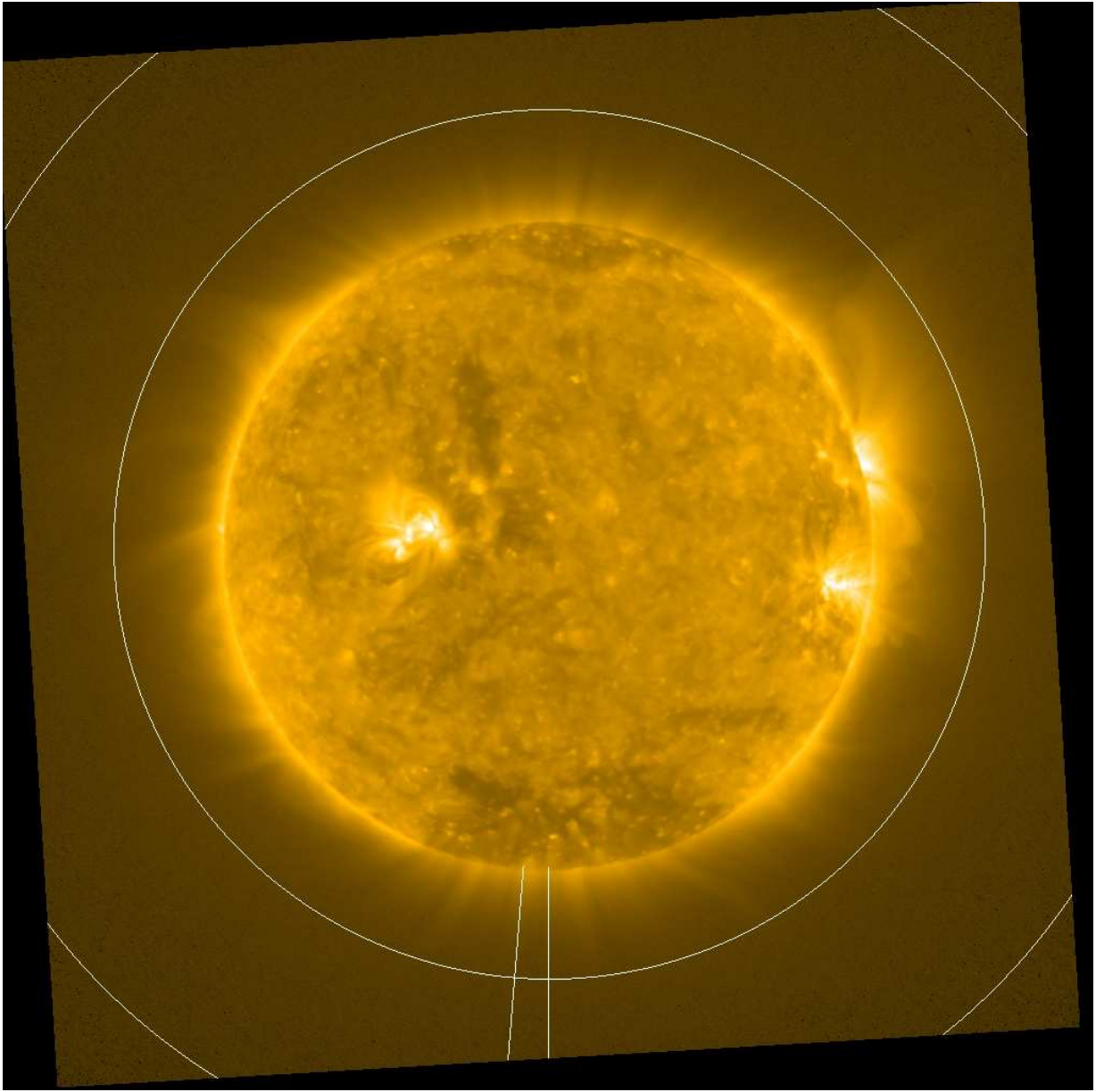}
	\caption{\label{fig:image} Image of the Sun from the SWAP data. Solar North is up. This is the first image in the dataset. The dark pixels at the edges indicate regions where the solar image fell outside of the field of view at any point during the observation period. Circles are drawn at 1.35~$R_{\sun}$ and 1.95~$R_{\sun}$. The inner radius of 1.35~$R_{\sun}$ represents the boundary within which our measurements are expected to be reliable. The outer radius of 1.95~$R_{\sun}$ is the inner boundary of the pixels that we used for quantifying the noise based on the analysis of the corners of the image \rev{(see Section~\ref{subsec:noise})}. The radial lines indicate the pixels chosen as representative of an interplume region (right) and a plume region (left; see also Figure~\ref{fig:charc}). 
}
\end{figure}

We studied data from SWAP taken over a 48 hour period from 2017-04-06 00:44 until 2017-04-07 23:59 UTC. Over this time there were 1415 images. SWAP takes images of the Sun in a bandpass centered at 174~\AA. The field of view is wide, though there is a relatively low spatial resolution of about 3.2$^{\prime\prime}$. The images are taken with a moderate, but irregular cadence ranging from 30~s to 35~min and with a median cadence of 110~s. All of these images have an exposure time of 10~s. Figure~\ref{fig:image} shows one image from the SWAP dataset. 

The data were calibrated using the standard SWAP data preparation routine, which centers and rotates the solar image, normalizes the data by exposure time, and subtracts dark current. The data preparation includes the option to perform a point spread function (PSF) correction to reduce the effect of scattered light. We mainly studied the raw data without this correction, but we do use the corrected data to quantify the stray light and associated noise sources, as discussed below in Section~\ref{subsec:noise}. 

We filtered the time series by removing frames that appear to contain a large number of hot pixels or cosmic rays that were not removed by the calibration routines. This was accomplished by requiring that intensity in a $200 \times 200$ pixel region in the corner of each image not exceed 10~$\mathrm{DN\,s^{-1}}$, using the SWAP units of data number per second. Such high count rates are unreasonable at large heights and only arise due to cosmic rays. This filtering removed 75 images, leaving 1340 for analysis.

\begin{figure}
	\centering \includegraphics[width=0.9\textwidth]{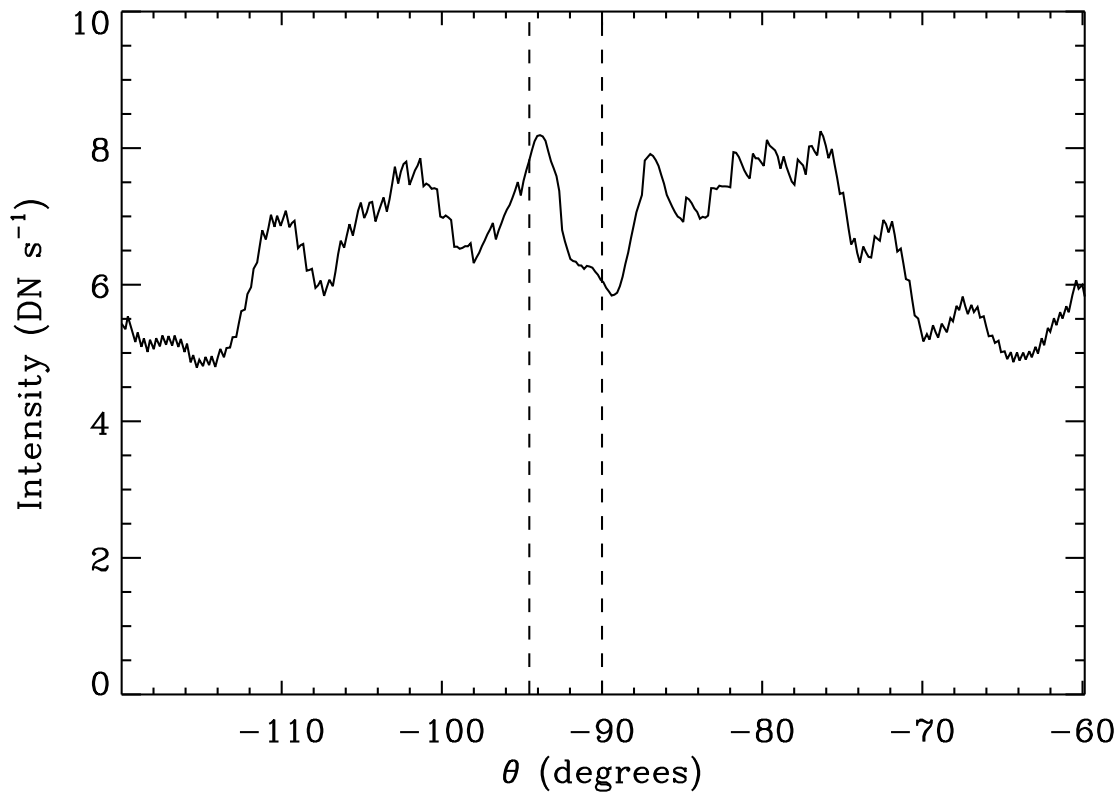}
	\caption{\label{fig:charc} Intensity at 1.1~$R_{\sun}$ as a function of angle in the coronal hole from the full time-averaged image. The interplume region we study lies along the merdian at $\theta = -90^{\circ}$, and the plume region is at $\theta \approx -95^{\circ}$, shown by the two dashed lines. The apparent noise in $I(\theta)$ moving away from $\theta = -90^{\circ}$ is due to the pixelization in the arc drawn at constant radius resulting in slight radial shifts by $\pm 1$ pixel between nearby angular locations. 
	}
\end{figure}

Figure~\ref{fig:charc} shows the intensity across the South polar coronal hole at a height of 1.1~$R_{\sun}$. We focus on a selection from an interplume and a plume region. The interplume region we selected lies along the central meridian at $\theta = -90^{\circ}$, while the plume region lies along the angle $\theta \approx -95^{\circ}$. Selecting these regions close to the meridian limits the influence of solar rotation on the observed variation. Nevertheless, we do see some evolution of these large scale structures over the 48~hr period of the observations, as will be discussed in more detail below. 

\section{Intensity Fluctuation Analysis}\label{sec:rms}

We have determined the root-mean-square (RMS) amplitude of the intensity fluctuations, $\Delta I$, in the corona as a function of height. One advantage of the RMS analysis is that it does not rely on a detailed evaluation of the power spectrum of the fluctuations, and so it can be performed up to larger heights where the noise level is greater, provided that the noise can be quantified and removed. This contrasts with previous analyses of intensity fluctuations, which have used Fourier or wavelet methods. But, because of the relatively high signal-to-noise ratio required for those analyses, the corresponding observations have tended to be performed close to the solar limb \citep[e.g.,][]{DeForest:ApJ:1998, Popescu:AA:2005, McIntosh:AA:2010}. A power spectrum type analysis is also complicated for the SWAP data because of the irregular cadence used in the observations. 

The RMS amplitude $\Delta f$ of a continuous signal, $f(t)$, is defined as 
\begin{equation}
\Delta f=\sqrt{ \frac{1}{T} \int_{T} f(t)^2 \diff{t} },
\label{eq:rmsdef}
\end{equation}
where $T$ is the total length over which the the signal is observed. If $f(t)$ is a sum of periodic functions at different frequencies, such as sine waves, and if each individual component has an RMS amplitude $\Delta f_{i}$, then the total measured RMS amplitude will be 
\begin{equation}
\Delta f_{\mathrm{signal}} = \sqrt{ \sum \Delta f_{i}^2 },
\label{eq:sumwaves}
\end{equation}
provided that the components are incoherent and the observation period is much longer than the period of any of the individual components. Thus, we can determine the RMS amplitude of intensity fluctuations in the coronal hole, but at the expense of information about the periods. Here and throughout all amplitudes are RMS amplitudes. 

\subsection{Quantifying Noise}\label{subsec:noise}

In addition to real signal, there are various noise sources that contribute to the measured amplitude. These noise sources add in quadrature with the real signal, so that the total measured $\Delta I_{\mathrm{meas}}$ can be described as 
\begin{equation}
\Delta I_{\mathrm{meas}}^2 = \Delta I^2 + \Delta I_{\mathrm{phot}}^2 + \Delta I_{\mathrm{det}}^2 + \Delta I_{\mathrm{scat}}^2. 
\label{eq:rmsnoise}
\end{equation}
Here, \rev{$\Delta I$ with no subscript} refers to real variations from density fluctuations in the corona, $\Delta I_{\mathrm{phot}}$ refers to the photon counting statistical noise, $\Delta I_{\mathrm{det}}$ is noise from spatial and time variations of the detector itself, and $\Delta I_{\mathrm{scat}}$ represents fluctuations in the scattered light intensity caused by real solar variations in other parts of the field of view, particularly on the solar disk. 

\subsubsection{Photon Counting Noise}

The photon noise follows Poisson statistics and can be estimated from the intensity itself. The SWAP calibration routines provide the intensity $I$ in $\mathrm{DN\,s^{-1}}$. The quantity that is actually counted is the charge and the SWAP inverse gain is $G=31$~$e^{-} \, \mathrm{DN}^{-1}$ \citep{Seaton:SolPhys:2013}. For the constant exposure time of $\Delta t=10$~s and a spatial binning over $N$ pixels, $\Delta I_{\mathrm{phot}}$ in units of $\mathrm{DN\,s^{-1}}$ is then given by 
\begin{equation}
\Delta I_{\mathrm{phot}} = \sqrt{ \frac{I} {N G \Delta t} }. 
\label{eq:rmsphot}
\end{equation}
For most of our analysis we use unbinned data with $N=1$. In order to determine $\Delta I_{\mathrm{phot}}$ for each pixel, we have used the intensity in each pixel averaged over the entire 48 hour observation period. It is expected that the extensive averaging removes other noise sources from the determination of $I$ for Equation~(\ref{eq:rmsphot}). Given the observed intensities, $R_{\mathrm{phot}}$ falls in the range of roughly 0.04--0.3~$\mathrm{DN\,s^{-1}}$. 

\subsubsection{Detector Noise}

The detector noise is the result of pixel-to-pixel variations in the detector sensitivity as well as the counting noise arising from the dark current. SWAP uses a CMOS-APS detector in which each pixel on the detector is read out separately and has its own set of electronics. The total detector noise level in the data convolves both a spatial noise and a temporal noise. The spatial variations arise because the solar image is not fixed with respect to the detector. Thus, each imaged spatial location in the field of view samples a large number of detector pixels and experiences pixel-to-pixel variations in the detector sensitivity. The temporal fluctuations occur because of the electronic dark current. The dark current causes what appears as an intensity background in the raw data. This intensity background is not caused by photons hitting the detector, but instead by currents in the electronics. This dark current contribution to the raw intensity is removed by the data preparation routines using a calibration for each detector pixel that is based on the temperature of the instrument as described by \citet{Halain:SolPhys:2013}. However, this correction applies only to the intensity itself and not to the variance of the intensity. The dark current causes noise in the intensity signal due to the counting noise associated with the dark current and uncertainties in the dark current correction. These residual fluctuations are not removed by the dark current correction. 

In order to estimate $\Delta I_{\mathrm{det}}$, we studied the RMS of intensity fluctuations at large heights above the Sun, $ > 1.95$~$R_{\sun}$, where there is expected to be little or no real emission from the corona and which we can use to infer the combined effects of the spatial and temporal noise of the detector. These heights correspond roughly to the corners of the detector. Because the pointing of the telescope is not fixed, the image moves around on the detector. During the data preparation, the images are rotated and co-aligned to all have the Sun at the center of the field of view with solar North up. This co-alignment results in some frames where pixels in the extreme corners of the data array correspond to locations on the sky that were not observed in every frame. Those pixels therefore have a value of zero intensity in the co-aligned data. We masked all such pixels out of our analysis. The total number of corner pixels at radius $r > 1.95$~$R_{\sun}$ that were never outside the field of view amount to a statistical sample of over 35,000 pixels. This sample of pixels is illustrated in Figure~\ref{fig:image} by the nonzero pixels that lie outside the circle at 1.95~$R_{\sun}$. 

We calculated the RMS of the intensity within each image for all the pixels meeting the above criteria and estimated $\Delta I_{\mathrm{det}}$ from the average over all the images. Based on that analysis we find $\Delta I_{\mathrm{det}} = 0.347 \pm 0.031$~$\mathrm{DN\,s^{-1}}$, where the uncertainty is the standard deviation of the RMS from each image over the entire set of images. This value of $\Delta I_{\mathrm{det}}$ is assumed to be constant across the detector. We have tested this assumption by performing the same analysis on various subsets of the corner pixels, such as only the top left corner, the bottom two corners, etc. and found that the various subsets give the same value of $\Delta I_{\mathrm{det}}$ to within the uncertainties. 

\subsubsection{Scattered Light}\label{subsubsec:scat}

In the corners, far from the Sun at $> 1.95$~$R_{\sun}$, there is still a non-zero intensity, which is likely dominated by scattered light. Scattered, or stray, light is caused by roughness of the mirror surface which scatters light from the solar disk to other parts of the detector \citep{Seaton:ApJ:2013}. Based on the intensity in the corners, averaged over all the frames, the scattered light level is very low, $I_{\mathrm{scat}} \approx 0.341 \pm 0.028$~$\mathrm{DN\,s^{-1}}$. Using Equation~(\ref{eq:rmsphot}), the expected photon noise from this level of intensity is only about $R_{\mathrm{phot}} \approx 0.03$~$\mathrm{DN\,s^{-1}}$, which is negligible compared to $\Delta I_{\mathrm{det}}$. Any photon noise associated with this stray light level is already accounted for in $\Delta I_{\mathrm{phot}}$, because that RMS estimate is based on the total intensity including both real emission and stray light. 


One reason that stray light may be important is that it can add noise due to the evolution of structures on the solar disk. These variations in the disk intensity are reflected as variations in the absolute scattered light intensity. To estimate the stray light fluctuations, we measured the average intensity of the corner pixels within each frame as a function of time. The stray light intensity in the corners is constant within each image, but can vary among the images due to the time variations of solar disk structures. The time variation of the average intensity in the corners and is found to be $\Delta I_{\mathrm{scat}}\approx 0.015$~$\mathrm{DN\,s^{-1}}$, or about 4.4\% of the absolute stray light intensity. This value should depend on the level of activity on the solar disk and is likely to vary among observations. \rev{Closer to the solar disk the RMS amplitude of these fluctuations should increase in proportion to the stray light intensity, $\Delta I_{\mathrm{scat}} = f_{\mathrm{scat}}I_{\mathrm{scat}}(x,y)$, where $I_{\mathrm{scat}}(x,y)$ is the absolute stray light intensity in each pixel at position $(x,y)$ and $f_{\mathrm{scat}}$ is a proportionality constant}. Based on the 4.4\% stray light fluctuation level in the corners, we set $f_{\mathrm{scat}}=0.044$.

\begin{figure}
	\centering \includegraphics[width=0.9\textwidth]{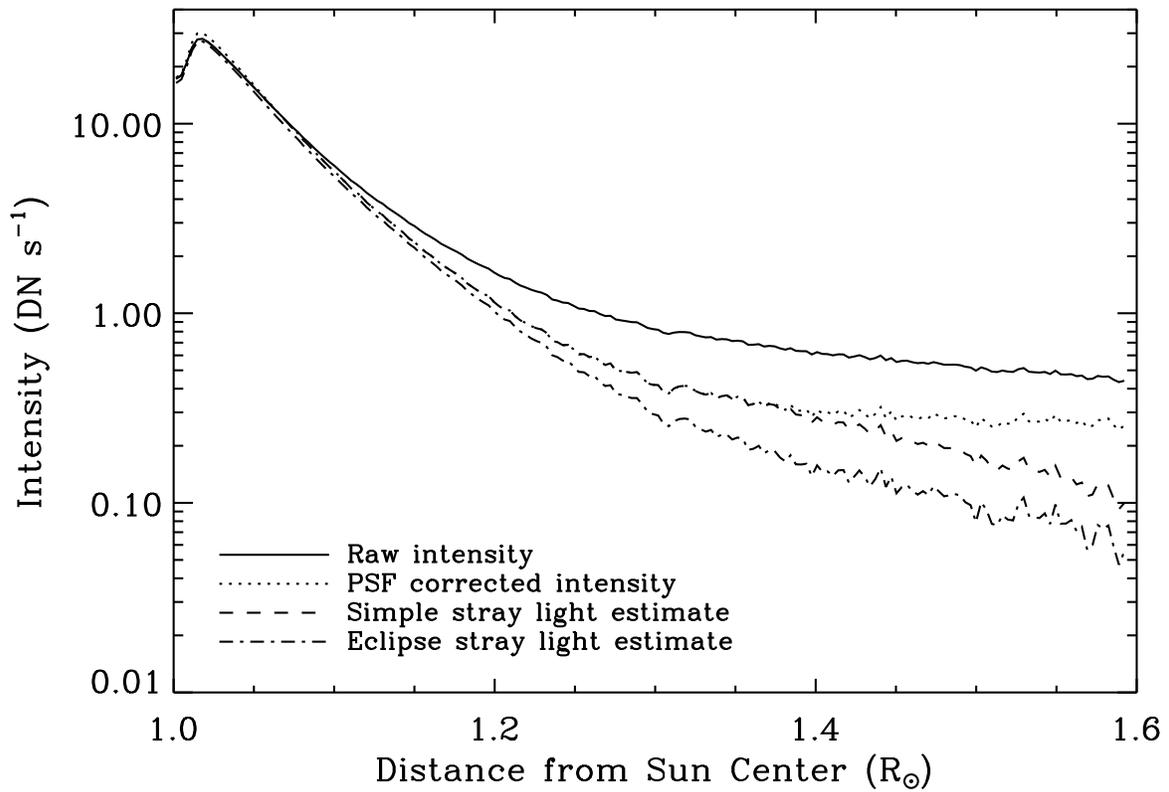}
	\caption{\label{fig:Itot} Average intensity as a function of radius in the interplume region. The solid curve shows the raw intensity and the dotted curve illustrates the effect of the PSF correction. This correction still leaves residual stray light at large heights. The other curves show intensity after removing an estimate of the stray light based on the PSF correction and the measured scattered light intensity in the corners (dashed) or using an estimate based on eclipse images (dash-dotted).
	}
\end{figure}

\begin{figure}
	\centering \includegraphics[width=0.9\textwidth]{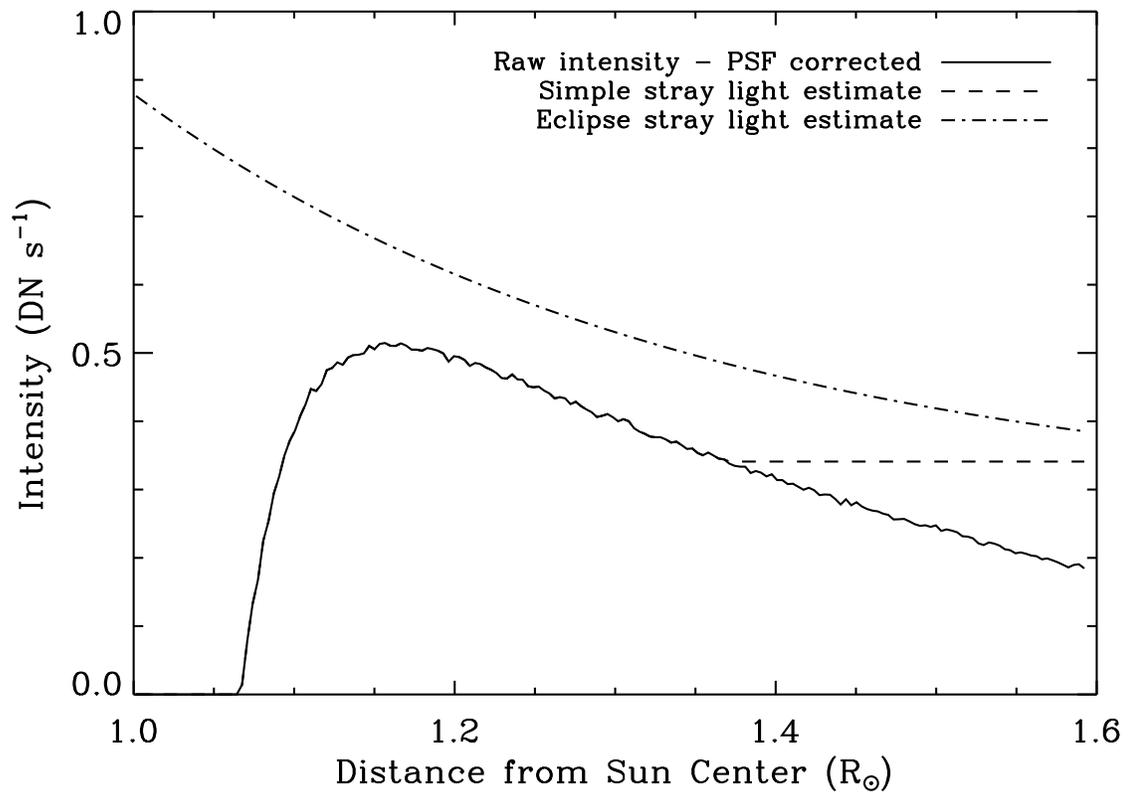}
	\caption{\label{fig:scat} Estimates of the stray light intensity as a function of radius along a radial slice in the interplume region. The solid curve shows the difference between the raw intensity and the PSF-corrected intensity. At large heights, the PSF correction does not remove all the stray light. The dashed curve indicates the stray light intensity based on the average intensity at very large heights. The dash-dotted curve is an alternative estimate of the scattered light based on eclipse images.
}
\end{figure}

Clearly, in order to account for the fluctuations of the scattered light, we need to know the stray light intensity as a function of position $I_{\mathrm{scat}}(x,y)$. We have used two methods to estimate the scattered light: 

First, we can patch together a stray light profile based on the point-spread function (PSF) correction and the stray light intensity measured in the corners of the image, described above. Near the disk the stray light can be removed by the data calibration using a point-spread function (PSF) correction \citep{Halain:SolPhys:2013, Seaton:ApJ:2013}. For our analysis we generally use the data without applying the PSF correction, since doing so could introduce systematic errors in our estimates of the noise. However, by subtracting the intensity with the PSF correction from the raw data we obtain an estimate of the stray light intensity in each pixel. Figure~\ref{fig:Itot} compares various estimates of the real intensity including the uncorrected total intensity and the intensity after applying the PSF correction. The PSF correction appears reasonable close to the disk, $\lesssim 1.2$~$R_{\sun}$, but at larger distances the images show evidence for residual stray light. \rev{The intensity is expected to decrease exponentially with distance. However, in Figure~\ref{fig:Itot} there is a clear flattening of $I(r)$ at large heights that indicates that the PSF-corrected intensity still contains residual stray light.} Thus, we may consider the PSF correction alone to give a lower bound on the stray light level (see Figure~\ref{fig:scat}).


To better remove this residual stray light, we assume a radially constant stray light level at large heights. Figure~\ref{fig:scat} illustrates several estimates of the scattered light at large heights. The dashed line in Figure~\ref{fig:scat} shows our best estimate, for which at large heights we take $I_{\mathrm{scat}}$ as either the value inferred by comparing the PSF correction to the raw data or the average $I_{\mathrm{scat}}$ inferred from the corners of the images ($I_{\mathrm{scat}}=0.341$~$\mathrm{DN\,s^{-1}}$), whichever is larger. Figure~\ref{fig:Itot} shows the effect of subtracting this stray light estimate from the raw intensity. 

\begin{figure}
	\centering \includegraphics[width=0.9\textwidth]{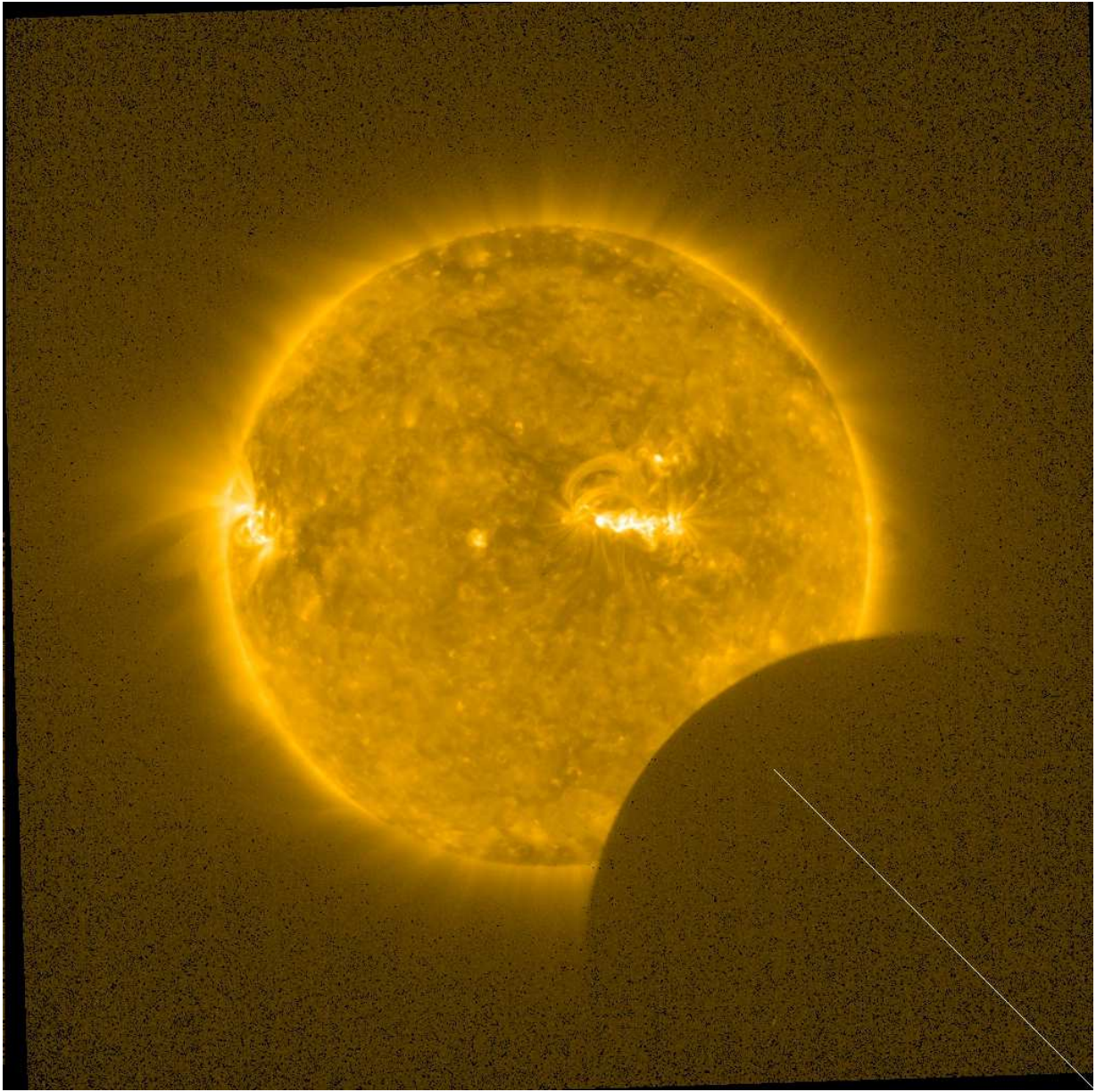}
	\caption{\label{fig:eclipse} SWAP image during the eclipse on 2017-08-21. The radial line illustrates the line along which the scattered light estimate was measured. 
}
\end{figure}

A second method for estimating the stray light is based on eclipse images, where part of the field of view is blocked by the moon. This method has previously been applied to SWAP data by \citet{Goryaev:ApJ:2014} using eclipse data from 2011-07-01. Here, we have used data from the solar eclipse on 2017-08-21, which occured only a few months after our main observational data were taken. Figure~\ref{fig:eclipse} shows one frame from our eclipse dataset. In this time period the moon is moving roughly downward and to the left through the images. We have compiled all the frames in which the radial line at $\theta=-45^{\circ}$ from the equator was covered by the Moon over the full extent from the solar limb to the edge of the field of view. Along this line, all of the intensity must be due to scattered light. 

\begin{figure}
	\centering \includegraphics[width=0.9\textwidth]{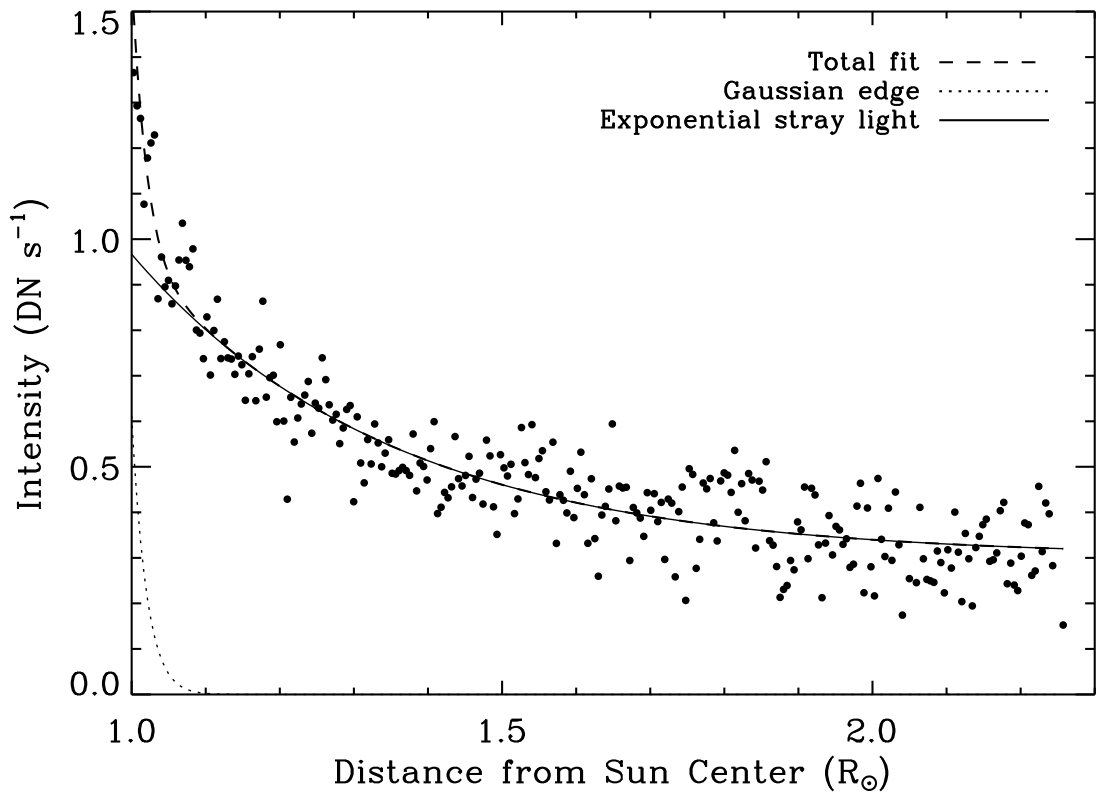}
	\caption{\label{fig:scateclipse} The intensity along the radial line highlighted in Figure~\ref{fig:eclipse} is indicated by the filled circles. The dashed line shows a fit to the data that is a sum of a Gaussian component (dotted curve) that describes the intensity affected by being at the edge of the moon during certain times of the observation and an exponential component (solid curve), which is ascribed to the stray light. Scaling this solid curve to match the intensity at large heights in our data gives the eclipse stray light estimate shown in Figure~\ref{fig:scat}.
	}
\end{figure}

Figure~\ref{fig:scateclipse} shows the intensity along the radial cut highlighted in Figure~\ref{fig:eclipse}. At large heights, the scattered light intensity is nearly constant with a value of about $I_{\mathrm{scat}} = 0.31$~$\mathrm{DN\,s^{-1}}$. Moving towards lower heights, the scattered light intensity slowly increases, and near the limb there is a sharper increase in the stray light profile. In some of the images used for the stray light analysis, the Moon barely covers the limb as it moves through the field of view during the exposure. So, the sharp increase in the intensity at these low heights is likely real emission. A similar profile was found by \citet{Goryaev:ApJ:2014}, who also ascribed the sharp increase near the limb to real emission. In order to quantify the stray light, we have fit the data in Figure~\ref{fig:scateclipse} with a sum of a Gaussian function and an exponential. The Gaussian part quantifies the emission at low heights, which we do not consider to be from the scattered light. The exponential part gives us a function describing the radial evolution of the stray light at large heights.

In order to apply this function to our analysis, we scale the inferred stray light profile to match the intensity at large heights. The resulting scattered light estimate is shown by the dot-dashed curve in Figure~\ref{fig:scat}. The effect of subtracting this stray light from the raw intensity is illustrated in Figure~\ref{fig:Itot}. The eclipse estimate gives a slightly larger value for the scattered light than the PSF correction. The eclipse estimate is expected to be more accurate at the large heights where the PSF correction breaks down, but near the limb the PSF correction is probably better. We use both estimates in our analysis, and consider the differences to be a systematic uncertainty. 

\subsection{RMS results}\label{subsec:rmsresults}

	We calculated the total RMS amplitdue of intensity fluctuations using, 
\begin{equation}
\Delta I_{\mathrm{meas}} = \sqrt{ \frac{1}{N} \sum_{j} \left[I_j(x,y)-\avg{I(x,y)} \right]^2 }. 
\label{eq:rmstotcalc}
\end{equation}
Here, $I_{j}(x,y)$ is the intensity in the pixel at position $(x,y)$ at the time labeled by index $j$ out of the $N$ total images. For the time-averaged intensity, $\avg{I(x,y)}$, we used two different schemes, both of which give qualitatively similar results. The most straightforward option is to take $\avg{I(x,y)}$ to be the average over the entire data set of 1340 frames. We will refer to fluctuations relative to this total average as the ``average-difference'' results. 

Another option is to calculate $\Delta I_{\mathrm{meas}}$ using a running difference. In this scheme, $\avg{I(x,y)}$ is the average over a certain number of frames previous to the current image, i.e., the average from $j-n$ to $j$. This method has often been applied to analysis of intensity fluctuations looking for periodic signals and the interval for the averaging $n$ is often chosen to be about $\sim20$~minutes \citep[e.g.,][]{Banerjee:AA:2001,Gupta:ApJ:2010}. Based on their results, we have chosen to average over $n=10$ frames in constructing our running-difference time series. Since these data have an average cadence of $110$~s this corresponds to an average interval of $18.3$ minutes. As shown below, in Section~\ref{subsec:periods}, the running difference suppresses contributions to $\Delta I_{\mathrm{meas}}$ from very low frequency changes, such as might be caused by slow variations in coronal structure or by solar rotation. 

\begin{figure}
	\centering \includegraphics[width=0.9\textwidth]{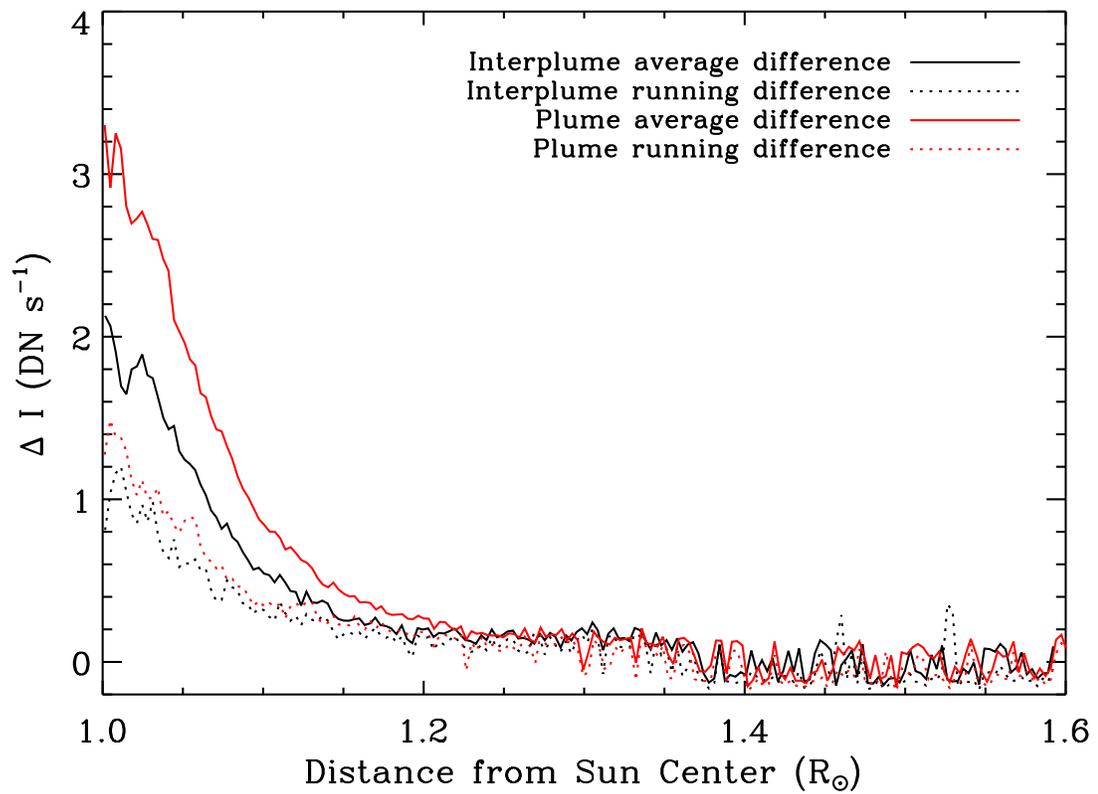}
	\caption{\label{fig:rmspip} RMS of intensity fluctuations, $\Delta I$, as a function of radius after correcting for noise. The black curves show the results for the interplume region and the red curves show the results for the plume. Solid curves indicate the RMS calculated using the average-difference method while the dotted curves were calculated using a running-difference method. See \rev{Section~\ref{subsec:rmsresults}} for details. 
}
\end{figure}

After calculating $\Delta I_{\mathrm{meas}}$ as a function of radius using both the average-difference and running-difference methods, we subtracted the noise contributions following Equation~(\ref{eq:rmsnoise}) in order to find $\Delta I$, the RMS due to real coronal fluctuations. Figure~\ref{fig:rmspip} shows $\Delta I$ as a function of radius in both the interplume and plume regions. In the figure the solid curves indicate $\Delta I$ computed using the average-difference method, while the dotted curves were computed using the running-difference method. It is clear that the choice of methods makes a significant difference at the lowest heights $\lesssim 1.2$~$R_{\sun}$ but above that height the differences are negligible. The changes in $\Delta I$ at low heights between the average- and running-difference methods are due to the suppression of low frequency fluctuations in the running-difference data, as shown below in Section~\ref{subsec:periods}. 

In all cases $\Delta I$ is approximately zero at the largest heights, which is evidence that the noise level is being subtracted correctly. Note that the $\Delta I$ calculated here is for a different subset of pixels than the subset that was used for the noise analysis based on the corners of the images. So, the fact that the $\Delta I$ goes to zero indicates that the noise parameters derived from the corners are still valid along the radial slices we have chosen for the analysis. 

In reality, there are likely to be real intensity fluctuations very far away from the Sun. That we find $\Delta I$ to be zero in our data is an artifact of our assumption that our data at large heights are dominated by noise. As a result we are not able to observe those real fluctuations. Hence, the fluctuations at the largest heights in our data are underestimated and we should limit our analysis to lower heights before $\Delta I$ goes to zero. For this reason, we discuss our results only for heights below about 1.35~$R_{\sun}$. This height is illustrated by the smaller circle in Figure~\ref{fig:image}. 

The physically significant quantity to be considered is $\Delta I$ normalized by the real intensity $I$. By real, we mean that $I$ here is the intensity after removing the contributions of scattered light. As the stray light level is uncertain, we use our estimate based on the PSF correction as well as our estimate based on eclipse images, as discussed in Section~\ref{subsubsec:scat}. Considering the influence on our relative RMS analysis: a larger value for the stray light corresponds to a smaller real intensity resulting in a larger $\Delta I/I$, whereas a smaller value for the stray light gives a larger real intensity and a smaller $\Delta I/I$. 

\begin{figure}
	\centering \includegraphics[width=0.9\textwidth]{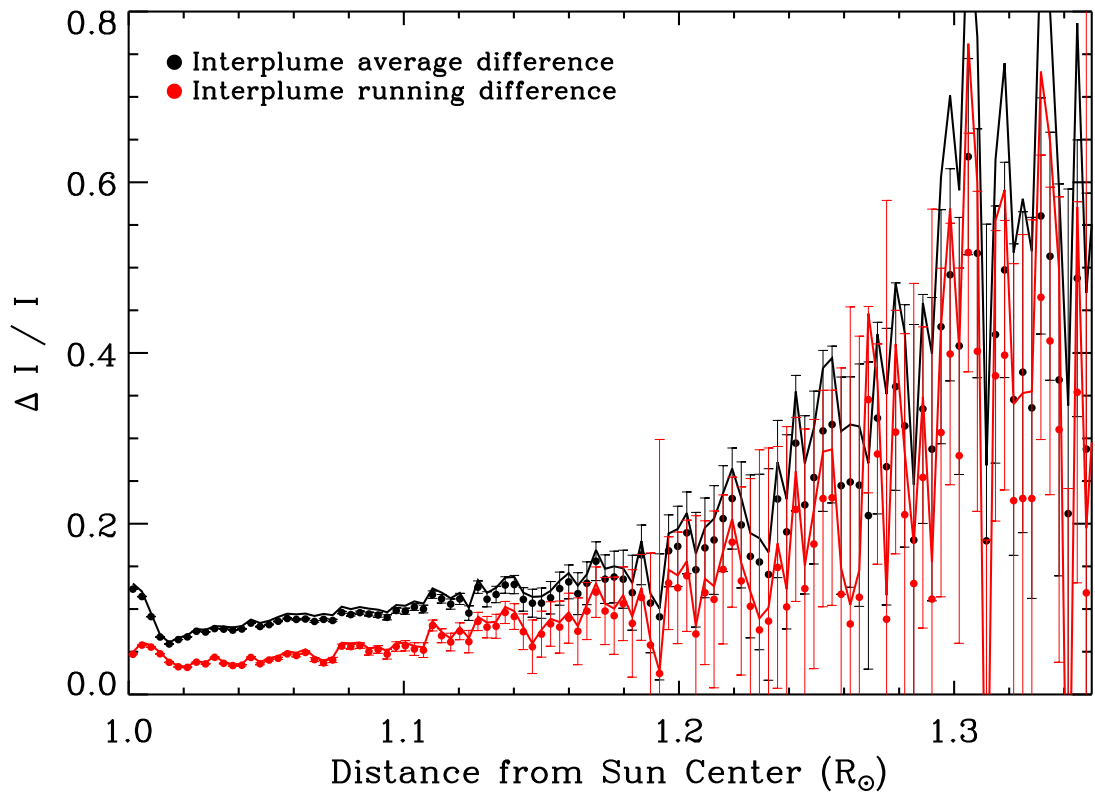}
	\caption{\label{fig:relrmsip} The RMS amplitude $\Delta I$ normalized by the intensity $I$ as a function of height in the interplume region. The black curve indicates results that were carried out using the average difference while the red curve shows the results for a running difference. For each case the filled circles show our results based on our PSF estimated stray light level. The error bars illustrate the statistical uncertainties due to the uncertainties in the various noise components of $\Delta I_{\mathrm{meas}}$. The solid curves indicate $\Delta I / I$ based on a normalization that uses the eclipse-based scattered light intensity.
	}
\end{figure}

\begin{figure}
	\centering \includegraphics[width=0.9\textwidth]{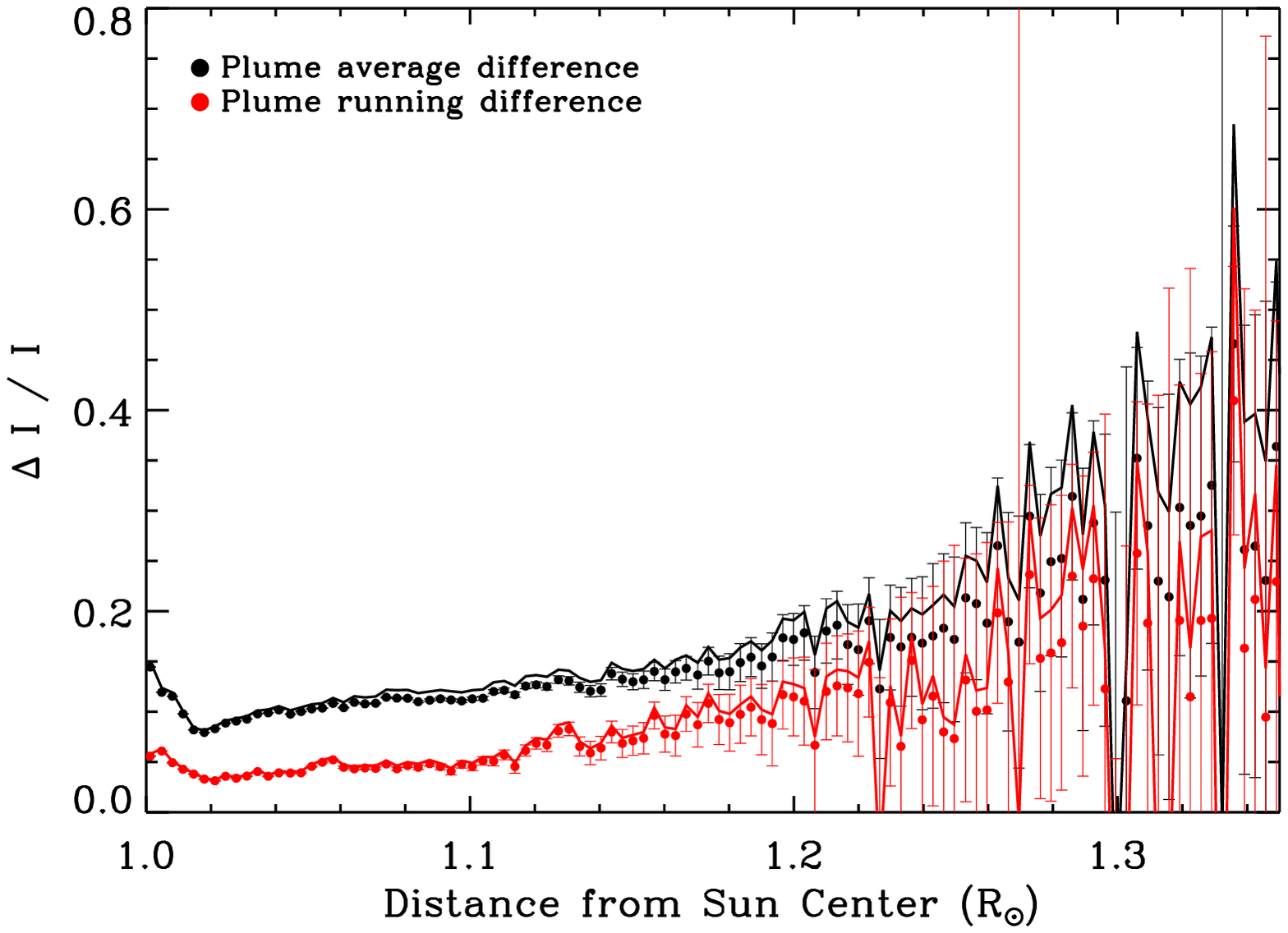}
	\caption{\label{fig:relrmspl} Same as Figure~\ref{fig:relrmsip}, but for the plume region.
}
\end{figure}

Figures~\ref{fig:relrmsip} and \ref{fig:relrmspl} show the inferred $\Delta I/I$ as a function of height in the interplume and plume regions respectively. In each plot, black points and curves indicate the analysis carried out using the average difference and red curves illustrate the results with the running difference. The filled circles indicate results using our PSF-based estimate for the stray light level. Solid lines indicate the results using the eclipse-based stray light level. The error bars are given only for the symbols and represent the propagated errors from the uncertainties in the various noise contributions to $\Delta I_{\mathrm{meas}}$. 

For both interplume and plume regions we see a very similar behavior. At low heights $\Delta I/I \sim 5$--$10 \%$, with the running-difference method giving a significantly lower estimate than the average-difference method. This is because at low heights the changing of background solar structures on long timescales is more prominent compared to higher frequency fluctuations, as we will discuss in the next section. Above about 1.15~$R_{\sun}$ the difference between the two analysis methods is no longer significant. 

\begin{figure}
	\centering \includegraphics[width=0.9\textwidth]{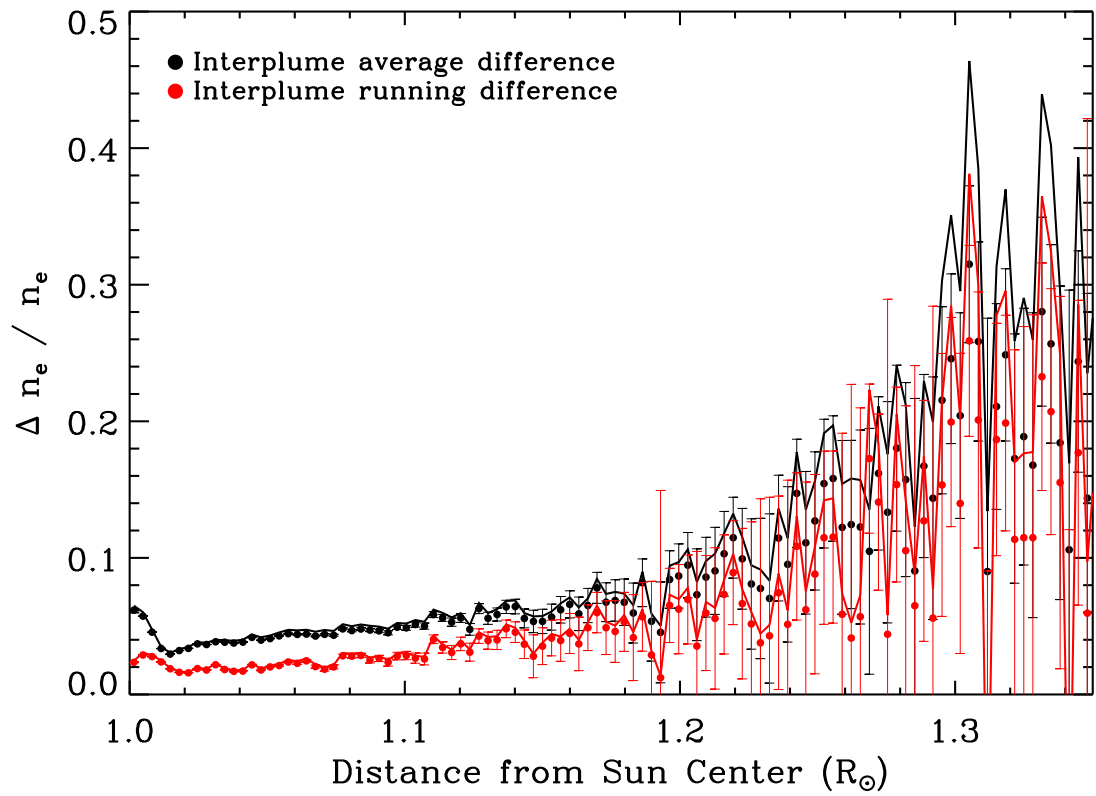}
	\caption{\label{fig:relrmsdenip} Same as Figure~\ref{fig:relrmsip}, but showing relative density fluctuations rather than relative intensity in the interplume region. 
}
\end{figure}

\begin{figure}
	\centering \includegraphics[width=0.9\textwidth]{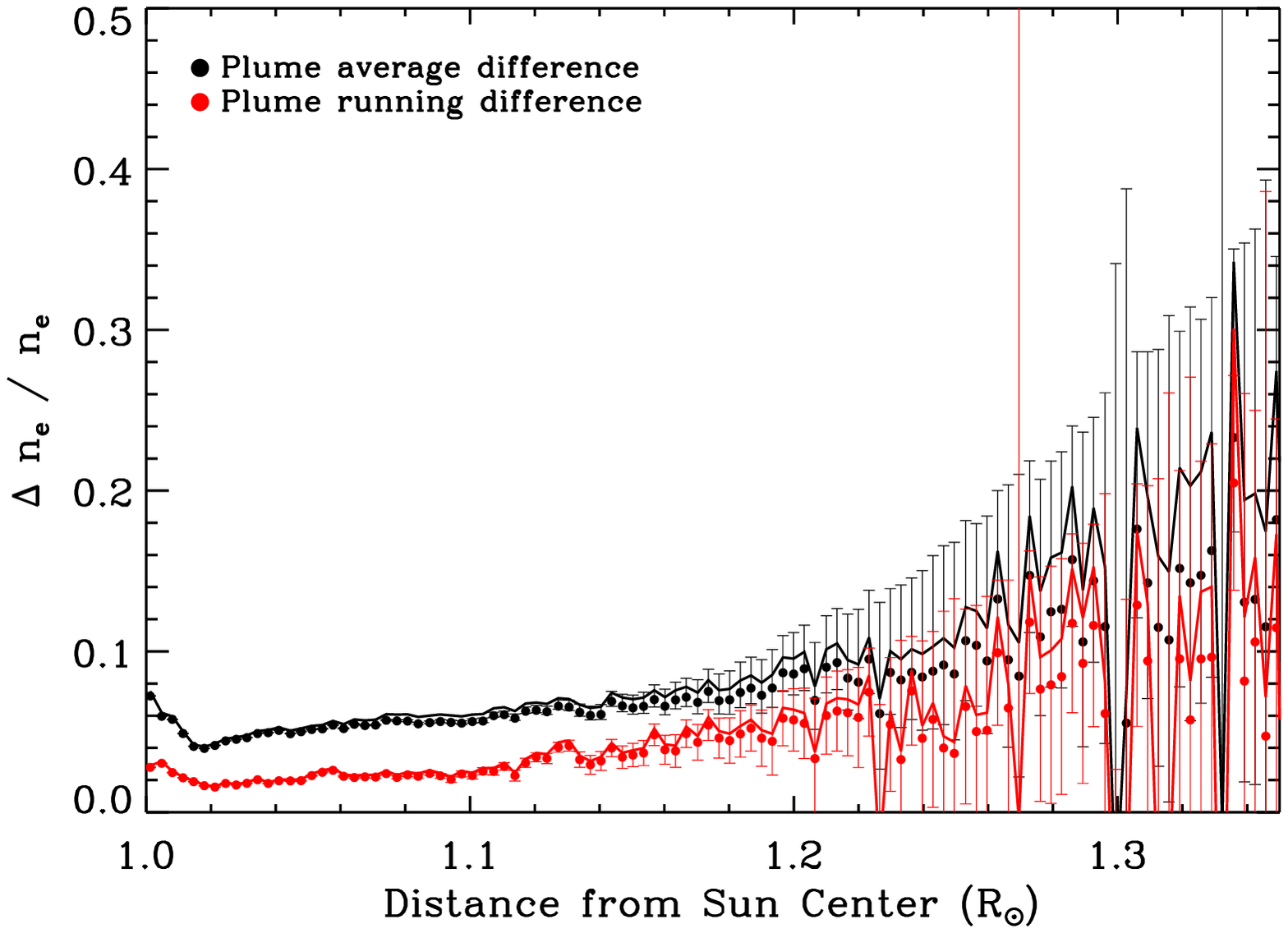}
	\caption{\label{fig:relrmsdenpl} Same as Figure~\ref{fig:relrmsdenip}, but for the plume region
}
\end{figure}

In all cases, regardless of the choice of background subtraction method, we see that there is an inflection point in $\Delta I/I$ between 1.1--1.2~$R_{\sun}$, where the relative intensity fluctuation amplitude begins to increase rapidly. Above 1.3~$R_{\sun}$, we find $\Delta I/I \sim 20$--$40\%$. This would correspond to a density fluctuation of $\Delta n_{\mathrm{e}}/n_{\mathrm{e}} \sim 10$--$20\%$, as illustrated graphically in Figures~\ref{fig:relrmsdenip} and \ref{fig:relrmsdenpl}. Although the observations include larger heights, the uncertainties on the noise sources and our assumption that the largest heights are completely dominated by noise preclude measurement of $\Delta I/I$ beyond about 1.35~$R_{\sun}$. Nevertheless from 1.1~$R_{\sun}$ up to 1.35~$R_{\sun}$ we find that there is a sharp increase in the relative RMS, indicating a corresponding increase in the amplitude of the underlying density fluctuations.

\section{Discussion}\label{sec:discuss}

\subsection{Periods}\label{subsec:periods}

We performed a periodogram analysis in order to understand the contribution of fluctuations at different frequencies to $\Delta I_{\mathrm{meas}}$. Because the intensity data were sampled at an uneven cadence, we used the Lomb-Scargle method to calculate the periodogram, $P(\omega)$ \citep{Scargle:ApJ:1982, Horne:ApJ:1986}. The periodogram gives the squared amplitude of the fluctuations as a function of angular frequency $\omega$. Analogous to Equation~(\ref{eq:sumwaves}) for a discrete set of wave amplitudes, the total RMS from $P(\omega)$ is the integral  
\begin{equation}
\Delta I_{\mathrm{meas}} = \sqrt{ \int P(\omega) \diff{\omega}}. 
\label{eq:rmsint}
\end{equation}

In calculating the periodogram, we have not attempted any correction for the various noise sources. Doing so would require an understanding not only of the RMS of the noise, but also of the frequency structure of the noise independent of any real fluctuations. Since we are not able to quantify such information in a reliable way, the periodogram is performed on the raw intensity data and therefore includes the noise. 

\begin{figure}
	\centering \includegraphics[width=0.9\textwidth]{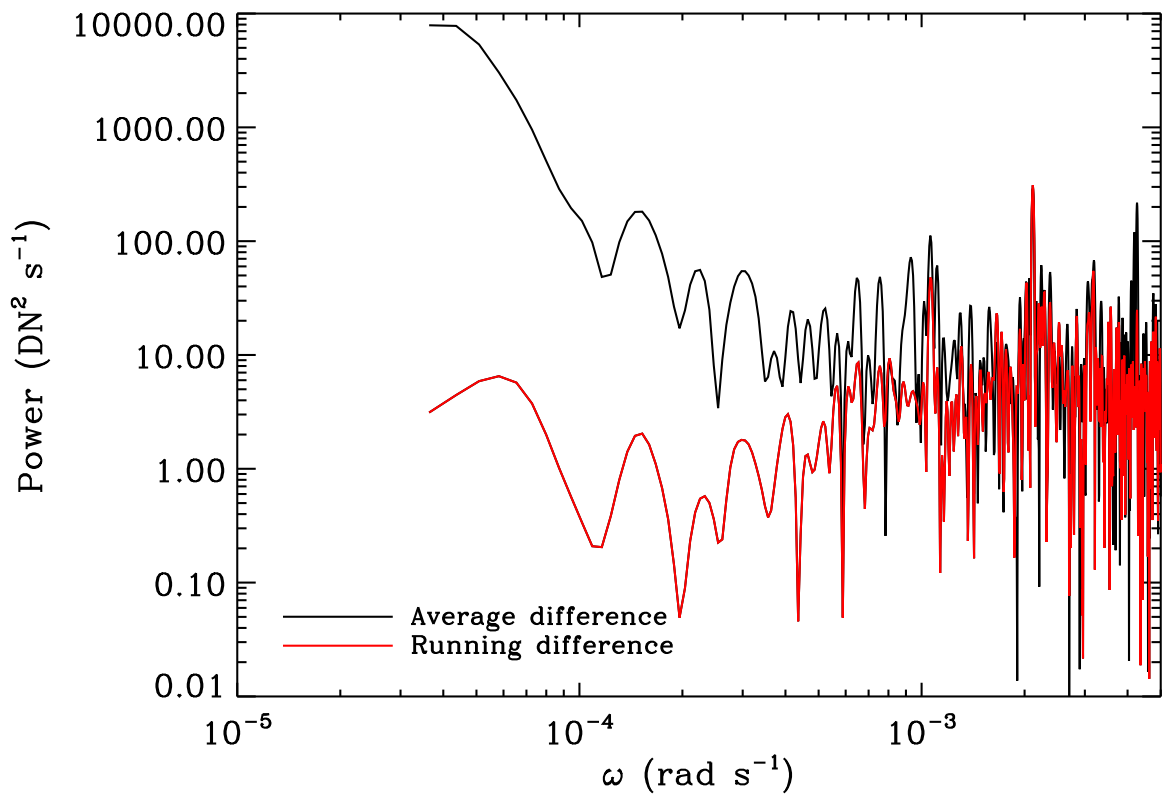}
	\caption{\label{fig:lowfreq} Periodogram for the interplume region at $r=1.05$~$R_{\sun}$ for low frequencies using both the average-difference (black curves) and running-difference (red curves) methods for calculating the fluctuations. Both curves show a peak at $\omega \approx 0.002$~$\mathrm{rad\,s^{-1}}$ corresponding to about a 50 minute period. The additional peak at $\omega = 0.004$~$\mathrm{rad\,s^{-1}}$ in the average-difference data is a harmonic of the $0.002$~$\mathrm{rad\,s^{-1}}$ frequency. 
}
\end{figure}

Figure~\ref{fig:lowfreq} shows the periodogram at $r=1.05$~$R_{\sun}$ in the interplume region at low frequencies and compares the average-difference to the running-difference method. This demonstrates that the main effect of the running difference is to suppress the low frequency fluctuations, as was mentioned above. Most of these fluctuations are likely to be due to long period changes in solar structure and to solar rotation. At higher frequencies the difference between the two methods is minor. Another minor change when comparing the periodogram with average versus running differences is that both methods show a strong periodicity at $\omega \approx 0.002$~$\mathrm{rad\,s^{-1}}$, but the analysis of the average-difference method also shows power at higher harmonics of this frequency, such as $\omega \approx 0.004$~$\mathrm{rad\,s^{-1}}$. In the running-difference analysis most of the power in these higher frequency harmonics is ascribed to the lowest frequency peak. Thus, the running-difference analysis seems to remove some ambiguity from the periodicity structure of the data. For the remainder of the discussion of the power spectrum, we focus on results using the running-difference method. We will also mainly describe the interplume region results, since the results for the plume region are qualitatively the same.

\begin{figure}
	\centering \includegraphics[width=0.9\textwidth]{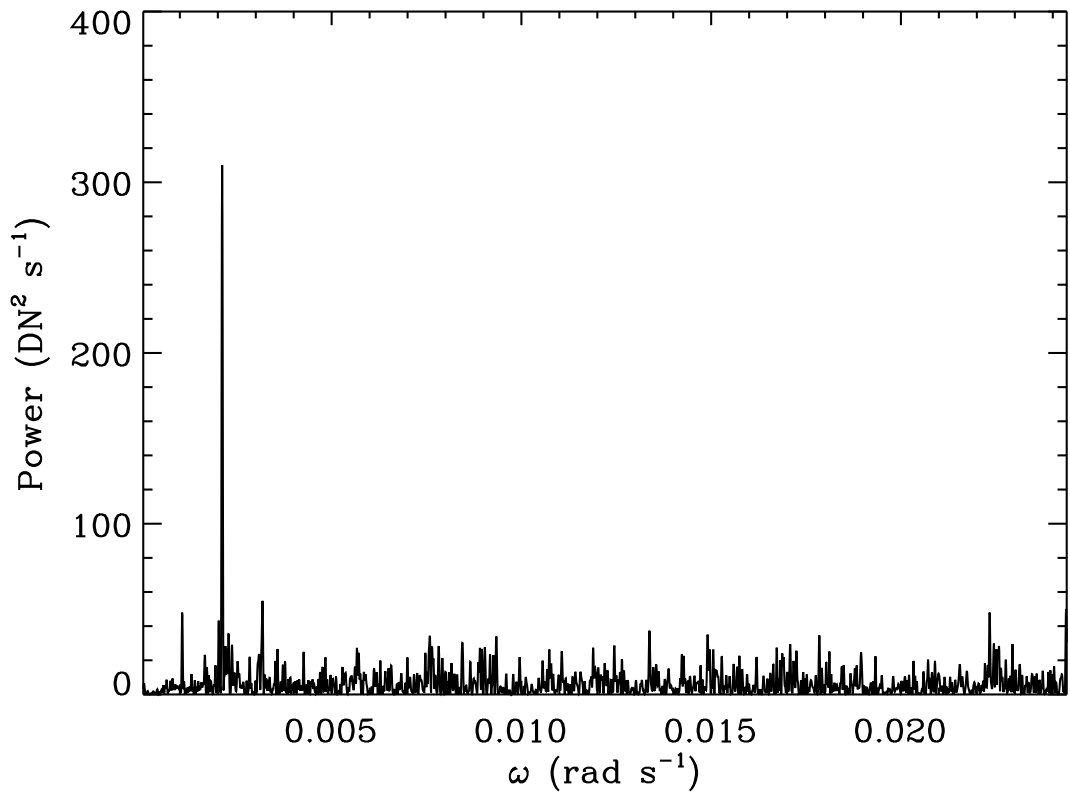}
	\caption{\label{fig:periodogram} Periodogram over the full frequency range for the interplume region at $r=1.05$~$R_{\sun}$ using the running-difference method. 
}
\end{figure}

The full periodogram for the interplume region at $r=1.05$~$R_{\sun}$ is shown in Figure~\ref{fig:periodogram}. One clear feature of the periodogram is the strong peak at $\omega = 0.0021$~$\mathrm{rad\,s^{-1}}$, corresponding to a period of about 50~min. It is possible that this is a real periodicty of acoustic waves in the coronal hole. Previous works looking at coronal holes have found periods of 10--30~min \citep{DeForest:ApJ:1998, Banerjee:SolPhys:2000, Popescu:AA:2005, Banerjee:AA:2009, Gupta:AA:2009, Gupta:ApJ:2010, Gupta:ApJ:2012, KrishnaPrasad:AA:2012, Liu:ApJ:2015},  as well as longer periods of 70--90 min \citep{Banerjee:AA:2001, Popescu:AA:2005}, and even up to 170 min \citep{Popescu:AA:2005}. However, it is suspicious that this period is very close to half the \textit{Proba2} orbital period of 98 min. Given the possibility that this period is a systematic effect, we quantify in detail its contribution to $\Delta I_{\mathrm{meas}}$, as is discussed below. 

\begin{figure}
	\centering \includegraphics[width=0.9\textwidth]{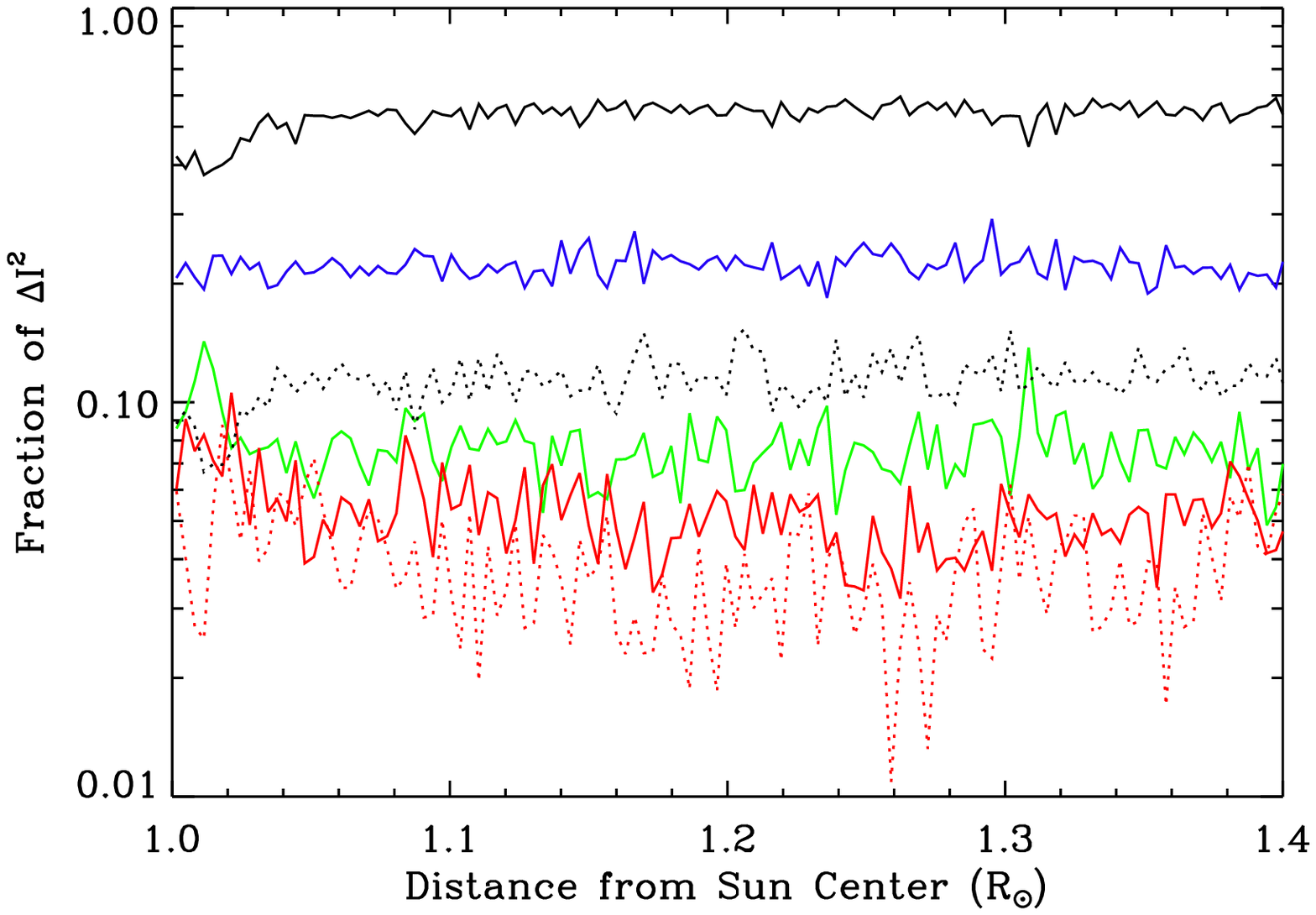}
	\caption{\label{fig:rmsperiods} \rev{The fraction of squared amplitude $\Delta I_{\mathrm{meas}}^2(\Delta \omega) / \Delta I_{\mathrm{meas}}^2$ contributed by various frequency or period intervals as a function of radius in the interplume region}. These were calculated by integrating portions of the periodogram \rev{(see Section~\ref{subsec:periods})}. The various intervals correspond to periods of 0--10~min (solid black curve), 10--20~min (solid blue curve), 20--30~mins (solid green curve), and 30--40~min (solid red curve). The dotted curves correspond to smaller intervals with special significance: the results for periods around 5 min are indicated by the dotted black curve and periods around 50 min by the dotted red curve.
}
\end{figure}

In order to understand the contribution of various periods to $\Delta I_{\mathrm{meas}}$ we integrated $P(\omega)$ over several intervals of $\omega$. These results are shown as a function of height in Figure~\ref{fig:rmsperiods}. Because the RMS components add in quadrature, we plot the fraction of the squared-RMS within each frequency interval, i.e., $\Delta I_{\mathrm{meas}}^2(\Delta \omega) / \Delta I_{\mathrm{meas}}^2$. These results show that the greatest contribution to the RMS comes from high frequencies corresponding to periods $T < 10$~mins and the next highest contribution is from periods in the range $T=10$--$20$~min. Longer periods in the ranges of $20$--$30$~min and $30$--$40$~min contribute less than 10\% to the total frequency-integrated $\Delta I_{\mathrm{meas}}^2$. It is also interesting that the high frequency, 0--10~min period, fraction of the RMS has a clear increasing trend between $1.0$--$1.1$~$R_{\sun}$, which implies that high frequency fluctuations are becoming more important relative to low frequency fluctuations in this height range. This implies that low frequency density fluctuations are being dissipated and/or that high frequency fluctuations are being excited.


Returning to the issue of the 50-min period, the red dotted curve in Figure~\ref{fig:rmsperiods} plots fraction of the $\Delta I_{\mathrm{meas}}^2$ from the region around this peak $\omega = 0.002$--$0.0022$~$\mathrm{rad\,s^{-1}}$ or $T = 47.6$--$52.4$~min. Despite the large magnitude of the peak in the periodogram, the contribution to the total fluctuation signal is only a few percent. It is also interesting that the relative amplitude of the 50-minute fluctuation has a decreasing trend with increasing height. One possibility is that the 50-minute period represents a long period acoustic wave that is excited at lower heights and dissipates. Regardless of the source of this periodicity, it is not an important contribution to $\Delta I_{\mathrm{meas}}$. 

Finally, a 5-minute periodicity is often expected for acoustic oscillations due to solar p-modes. In order to see the effect of signals with similar periods in our data, we calculated the RMS by integrating $P(\omega)$ over the interval $\omega = 0.020$--$0.023$~$\mathrm{rad\,s^{-1}}$ or $T=4.6$--$5.2$~min. These results are shown by the black dotted curve in Figure~\ref{fig:rmsperiods}, which indicates that fluctuations with periods near 5 min account for about 10\% of the total intensity fluctuation. 

\subsection{Energy Flux}\label{subsec:propagation}

If we assume that the density fluctuations represent linear acoustic magnetohydrodynamic wave modes, then the energy flux of the waves can be estimated on the basis of the perturbed velocity $\Delta v$, which is related to $\Delta I$ by \citep[see for example][]{Gurnett:book}
\begin{equation}
\Delta I /I = 2 \Delta v/ c_{\mathrm{s}}, 
\label{eq:dv}
\end{equation}
where $c_{\mathrm{s}}$ is the sound speed. The energy flux is then 
\begin{equation}
F \approx \frac{1}{2} \rho (\Delta v)^2 c_{\mathrm{s}}, 
\label{eq:eflux}
\end{equation}
with $\rho \approx m_{\mathrm{p}} n_{\mathrm{e}}$ the mass density. This estimate ignores the solar wind flow velocity, which is expected to be small at these low heights \citep{Cranmer:ApJ:1999}. 

Previous observations have suggested that the density fluctuations are sound waves propagating at the sound speed, at least for the fluctuations near the limb where the signal is strongest. Such observations have shown that near the limb the density fluctuations propagate radially outward with speeds of 75--150 $\mathrm{km\,s^{-1}}$, which is consistent with the sound speed in the corona \citep{DeForest:ApJ:1998, Banerjee:AA:2009, Gupta:AA:2009,Gupta:ApJ:2010, Banerjee:SSR:2011, Liu:ApJ:2015, McIntosh:AA:2010}. 

Our data are consistent with a propagation speed similar to these at low heights. However, because of the cadence and spatial resolution in our data, we were not able to obtain a more precise propagation speed than this estimate. For a speed of $100$-$150$~$\mathrm{km\,s^{-1}}$ and the $\sim 3^{\prime\prime}$ pixel size, the pixel crossing time is about 15--20~s. At a median cadence of 110~s, an intensity fluctuation travels vertically through at least 5--8 pixels from one frame to another. Due to the decrease in the absolute fluctuation amplitudes near the limb, the decrease in intensity with height, and the rather slow cadence there is a large uncertainty in the propagation speed. 
The analysis is further complicated by the irregularity of the cadence. Correlation methods for irregularly spaced data can be performed \citep[e.g.,][]{Edelson:ApJ:1988}, but the resolution is still limited.

In order to estimate the energy flux, we take $c_{\mathrm{s}}=150$~$\mathrm{km\,s^{-1}}$, corresponding to a temperature of 1~MK. A typical coronal hole density at 1.3~$R_{\sun}$ is $n_{\mathrm{e}} \approx 10^{7}$~$\mathrm{cm^{-3}}$. At that height, we find $\Delta I/I \approx 0.4$. Equations~(\ref{eq:dv}) and (\ref{eq:eflux}) then give $F \approx 10^{3}$~$\mathrm{erg\,cm^{-2}\,s^{-1}}$. The geometrical area expansion factor $A(r)/A(R_{\sun})$ for a flux tube in a coronal hole at 1.3~$R_{\sun}$ is about 3.2 \citep{Cranmer:ApJ:1999}. Therefore, this suggests that the required input energy flux at the base of the corona is at least $3.2 \times 10^{3}$~$\mathrm{erg\,sm^{-2}\,s^{-1}}$. Due to dissipation, the actual energy flux required at the base of the corona to support the observed density fluctuations at 1.3~$R_{\sun}$ is likely to be larger.

\subsection{Implications}\label{subsec:future}

One of the most interesting results shown by these data is that the relative amplitude of the density fluctuations increases with height, especially above about $1.2$~$R_{\sun}$. These results are roughly consistent with the few other observations of density fluctuations at large heights. For example, \citet{Mancuso:AA:2016} studied an O~\textsc{vi} line observed by UVCS on \textit{SOHO} and found $\Delta I/I \approx 10\%$ at 1.4~$R_{\sun}$. Although this is somewhat smaller than our estimate of $20$--$40\%$, it is within the uncertainties. 

At larger heights from about 1.5--6~$R_{\sun}$, \citet{Miyamoto:ApJ:2014} used radio occulation measurements to observe compressive waves with amplitudes $\Delta n_{\mathrm{e}}/n_{\mathrm{e}}$ growing from $\approx 1\%$ to $\approx 30\%$ over their observed height range. Those measurements may not be directly comparable to ours as the observations were taken in 2011 when the Sun was active and there was not a clear coronal hole near the poles. So those data appear to be from closed field regions. But, based on our measurements and all these earlier results it seems established that density fluctuations grow with height in the corona.

The observed density fluctuations cannot be the results of acoustic waves excited at the base of the corona that grow in amplitude as they propagate outward. Such waves are expected to be damped rapidly in the corona \citep{Ofman:ApJ:1999, Ofman:ApJ:2000}. Futhermore, even if they were undamped their amplitudes should be reduced by the geometric expansion of the magnetic flux tubes with radius \citep{DeMoortel:AA:2003, DeMoortel:AA:2004}. Previous observations have shown that density oscillations are rapidly dissipated near the limb of the Sun \citep[e.g.,][]{Gupta:AA:2014}. This conclusion is also supported by our period analysis showing that the relative contribution of high frequency fluctuations increases with height, which suggests that low frequency fluctuations are damped and/or that high frequency fluctuations are excited in the corona.

One explanation for the generation of density fluctuations that we find between 1--1.35~$R_{\sun}$ is that they are excited by the Alfv\'en waves through the parametric instability. The parametric instability is a wave-wave interaction in which an outward traveling finite amplitude Alfv\'en wave generates a longitudinal compressive wave as well as forward and backward propagating transverse magnetic waves \citep{Goldstein:ApJ:1978, LDelZanna:AA:2001}. Some models for wave heating of the corona have considered this mechanism as a way to dissipate energy carried by Alfv\'en waves into heat in the corona \citep{Suzuki:ApJ:2005, Shoda:ApJ:2018}. The recent model of \citet{Shoda:ApJ:2018} predicts density fluctuations that grow to 20\% by 1.4~$R_{\sun}$, depending on the transverse correlation length parameter used in their model. These values are in reasonable agreement with what we observe. Additionally, the generation of sunward propagating Alfv\'en waves by the parametric instability and the inward reflection of outward propagating waves from the density fluctuations is predicted to drive turbulence and result in heating of the corona \citep{vanB2016, vanB2017}. 


Our previous spectroscopic observations of emission line widths have suggested that Alfv\'en wave amplitudes are consistent with energy conservation up to about 1.1~$R_{\sun}$ and dissipation above that height \citep{Hahn:ApJ:2012, Hahn:ApJ:2013a}. It is interesting that the density fluctuations found here appear to show an inflection point at a similar height range, above which the amplitude of the fluctuations increases rapidly. This suggests that the apparent damping of the Alfv\'en waves is related to the increasing amplitude of the density fluctuations. 

Our spectroscopic observations implied that the dissipated Alfv\'en wave energy flux is on the order of a few times $10^{5}$~$\mathrm{erg\,cm^{-2}\,s^{-1}}$ at 1~$R_{\sun}$, after accounting for the geometric expansion. This is significantly larger than the density fluctuation energy flux of $\sim 3\times 10^{3}$~$\mathrm{erg\,cm^{-2}\,s^{-1}}$ estimated here. But, this estimate does not take into account the dissipation of the density fluctuations. More detailed calculations taking into account such sinks of energy are needed to determine precisely the required Alfv\'en wave energy that must be dissipated to produce the observed density fluctuations. For the present, though, the energy flux of the Alfv\'en waves appears to be sufficient to generate the observed density fluctuations. 

\section{Conclusions}\label{sec:conclusions}

We have studied intensity fluctuations at moderate heights from about $1$--$1.35$~$R_{\sun}$ observed with the SWAP instrument on \textit{Proba2}. By measuring the amplitude of the fluctuations based on the RMS and accounting for various systematic sources of noise, we were able to determine the RMS amplitudes of coronal intensity fluctuations to heights in the corona where the signal becomes weak. These intensity fluctuations are proportional to density fluctuations in the corona. We find that the relative density fluctuation amplitude, $\Delta n_{\mathrm{e}}/n_{\mathrm{e}}$, increases with height, with a rapid increase beginning between 1.1--1.2~$R_{\sun}$. These measurements imply that the density fluctuations are generated in the corona, rather than propagating upward from lower heights. A possible explanation for their generation is that they are excited by a parametric instability of Alfv\'en waves. This mechanism is consistent with both theoretical models of Alfv\'en wave heating in the corona \citep[e.g.,][]{Suzuki:ApJ:2005, Shoda:ApJ:2018} and previous observations \citep{Hahn:ApJ:2012, Hahn:ApJ:2013a}, suggesting that Alfv\'en waves begin to dissipate at heights in the corona similar to those where we find an increase in the density fluctuations. The presence of density fluctuations is also predicted to drive Alfv\'en wave reflection, leading to turbulence and coronal heating \citep[e.g.,][]{vanB2016, vanB2017}. 

Altogether these observations and theories support a model for heating coronal holes in which Alfv\'en waves are generated at the base of the corona. They propagate upward and undergo a parametric instability that generates density fluctuations and other magnetic waves. Sunward propagating Alfv\'en waves are generated by the parametric instability and by reflection off the density fluctuations. The nonlinear interaction of the counter-propagating Alfv\'en waves leads to turbulence. Finally, heating occurs both through the dissipation of the density fluctuations and due to turbulent heating.

\begin{acknowledgments}
	We thank Daniel Seaton, Mahboubeh Asgari-Targhi, and Sayak Bose for helpful discussions. M.H. and D.W.S were supported in part by the NASA Living with a Star Program grant NNX15AB71G and by the NSF Division of Atmospheric and Geospace Sciences SHINE program grant AGS-1459247. E.D. acknowledges support from the Belgian Federal Science Policy Office (BELSPO) through the ESA-PRODEX programme, grant No.\ 4000120800. SWAP is a project of the Centre Spatial de Li\'ege and the Royal Observatory of Belgium funded by BELSPO.
\end{acknowledgments}

\newpage
\clearpage

\clearpage

\bibliography{acoustic}

\end{document}